\documentclass[twocolumn,iop]{emulateapj}

\usepackage{amsmath,amssymb}
\usepackage{apjfonts}
\usepackage{mathrsfs}
\usepackage{bm}
\usepackage{color}

\newcommand{\beq}{\begin{equation}}
\newcommand{\eeq}{\end{equation}}

\newcommand{\pdot}{\ensuremath{\dot{P}}}
\newcommand{\ttf}{\ensuremath{t_{\rm TF}}}
\newcommand{\tv}{\ensuremath{t_{\rm V}}}

\newcommand{\msun}{\ensuremath{M_\odot}}
\newcommand{\mdot}{\ensuremath{\dot{M}}}
\newcommand{\mout}{\ensuremath{\mdot_{\rm out}}}
\newcommand{\vesc}{\ensuremath{v_{\rm esc}}}
\newcommand{\rtau}{\ensuremath{R_{\tau = 1}}}
\newcommand{\actaa}{Acta Astronomica}

\shorttitle{Burying a binary: V1309 Sco}
\shortauthors{O. Pejcha}

\begin{document}

\title{Burying a Binary: Dynamical Mass Loss and a Continuous Optically-Thick Outflow Explain the Candidate Stellar Merger V1309 Scorpii}

\author{ Ond\v{r}ej Pejcha\altaffilmark{1} }
\affil{Department of Astronomy, The Ohio State University, 140 West 18th Avenue, Columbus, OH 43210, USA}
\affil{Department of Astrophysical Sciences, Princeton University, 4 Ivy Lane, Princeton, NJ 08540, USA}
\altaffiltext{1}{Hubble and Lyman Spitzer Jr. Fellow}
\email{pejcha@astro.princeton.edu}

\begin{abstract}
V1309~Sco was proposed to be a stellar merger and a common envelope transient based on the pre-outburst light curve of a contact eclipsing binary with a rapidly decaying orbital period. Using published data, I show that the period decay timescale $P/\pdot$ of V1309~Sco decreased from $\sim 1000$ to $\sim 170$\,years in $\lesssim 6$\,years, which implies a very high value of $\ddot{P}$. I argue that V1309~Sco experienced an onset of dynamical mass loss through the outer Lagrange point, which eventually obscured the binary. The photosphere of the resulting continuous optically-thick outflow expands as the mass-loss rate increases, explaining the $\sim 200$\,day rise to optical maximum. The model yields the mass-loss rate of the binary star as a function of time and fits the observed light curve remarkably well. It is also possible to observationally constrain the properties of the surface layers undergoing the dynamical mass loss. V1309~Sco is thus a prototype of a new class of stellar transients distinguished by a slow rise to optical maximum that are driven by dynamical mass loss from a binary. I discuss implications of these findings for stellar transients and other suggested common envelope events. 
\end{abstract}

\section{Introduction}

V1309~Scorpii developed an unusually red color during its 2008 outburst and joined the group of enigmatic objects such as V838~Mon, V4332~Sgr, and M31 RV \citep[e.g.][]{martini99,munari02,mason10}.  One possible explanation of these objects is that they are mergers of stars \citep[e.g.][]{soker03,soker06,tylenda06}. V1309~Sco solidifies this view, because the pre-outburst photometry revealed a binary star system with a rapidly decaying orbital period (\citealt{tylenda11}; Fig.~\ref{fig:period}).

To make two stars in a binary merge, secular and dynamical instabilities must bring them close enough so that they orbit in a non-corotating common envelope, where the associated drag forces causes the final merger and envelope ejection \citep[e.g.][]{paczynski76,livio88,iben93,taam00,ivanova13b}.
V1309~Sco offers an excellent opportunity to robustly study the approach to merger through its orbital period decay and pre-outburst light curve.

In this paper, I focus on two important but neglected features of V1309~Sco: (1) the rapidly accelerating orbital period decay and (2) the slow $\sim 200$\,day rise to optical maximum. In Section~\ref{sec:period}, I argue that the orbital period decay is due to dynamical mass loss. In Section~\ref{sec:wind}, I illustrate how the the dynamical mass loss from the binary could yield to a slow brightness increase. In Section~\ref{sec:appl}, I present a fit to the light curve and period evolution of V1309~Sco and interpret the fit parameters within the dynamical mass loss model. In Section~\ref{sec:disc}, I discuss my findings and provide implications for other transients. In Section~\ref{sec:conc}, I review the results.

\section{Period change of V1309 Sco}
\label{sec:period}

\begin{figure}
\plotone{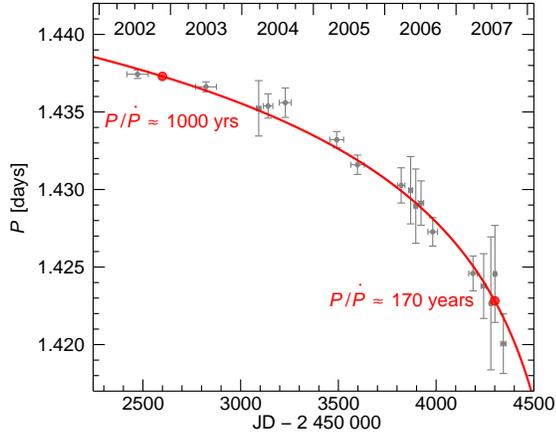}
\caption{Period evolution of V1309~Sco. Solid circles with errorbars show the period measurements of \citet{tylenda11}. The red line is a model combining the period and light curve measurements (Eq.~[\ref{eq:pdotp}], Fig.~\ref{fig:lc}); the implied mass loss rate is shown in Figure~\ref{fig:mdot}. The rapidly decreasing period decay timescale $P/\pdot$ is evaluated at two positions (red solid circles).}
\label{fig:period}
\end{figure}

Figure~\ref{fig:period} shows the orbital period evolution of V1309~Sco prior to its outburst as measured by \citet{tylenda11}. The period decrease, likely the fastest ever observed in a binary system, is accelerating with time. The decay timescale, $P/\pdot$, decreased from $\sim 1000$\,years to $\sim 170$\,years in less than $6$ years. During the same time, the second derivative timescale $\pdot/\ddot{P}$ decreased from  about $5.2$\,years to $1.1$\,years, suggesting that the period decay accelerated and higher-order derivatives of the period were important as well. A change in $P/\pdot$ by a factor of $\sim 6$ over a timescale that is at least order of magnitude shorter than $P/\pdot$ is very surprising. Since the period decrease eventually ended in an outburst and potentially a merger, it is imperative to understand the mechanism of the period change.

Understanding the observed period change is not straightforward. For example, one of the possible explanations of the period decay in V1309 Sco is the Darwin instability \citep{stepien11}, possibly in combination with other effects such as the mass loss from L2 \citep{tylenda11}. In the theory of equilibrium tides, the Darwin instability proceeds on the tidal friction timescale $\ttf$. Before the outburst, V1309~Sco was a contact binary filling its Roche lobes, which gives
\beq
\ttf \approx 80\tv \frac{(1+q)^{0.6}}{q^{0.64}}(1-Q_{\rm E})^2,
\label{eq:ttf}
\eeq
where $\tv$ is the intrinsic viscous timescale of the star, $Q_{\rm E}$ measures the quadrupolar deformability of the star\footnote{The coefficient $Q_{\rm E}$ is denoted as $Q$ in \citet{eggleton98} and \citet{eggleton01}, but it is a different quantity than the tidal quality factor also usually denoted as $Q$.} and $q=M_2/M_1$  is the mass ratio with $0.1 \lesssim q \lesssim 10$ for this approximation \citep{eggleton83,eggleton01}. Since $\tv$ is typically estimated to be between about a year and decades and $Q_{\rm E} \sim 0.2$ for a fully convective star \citep{eggleton98}, $\ttf$ is between about a hundred and thousands of years. The Darwin instability is driven by the subsynchronous rotation of one of the stars, $P/\pdot \sim \ttf (1-\Omega/\omega)^{-1}$, where $\Omega/\omega$ is the ratio of stellar spin frequency to the orbital frequency. The period decay is thus much slower for nearly synchronous rotation. The structure of the orbital evolution equations of \citet{eggleton01} suggests that the second derivative of period change is $\pdot/\ddot{P} \sim \ttf$ and that for any tidal process\footnote{Orbital braking mechanisms due to a magnetized stellar wind also affect the orbit on a timescale longer than $\ttf$ since the spin angular momentum loss of the star is transferred to the orbit by tidal forces.} both $\pdot$ and $\ddot{P}$ change on the tidal timescale $\ttf$.  However, the observations of the period change show that $P/\pdot$ changes from about $1000$ to $170$\,years and $\pdot/\ddot{P}$ from about $5$\,years to $1$\,year, which is noticeably shorter than $\ttf$. These two observed timescales differ by about two orders of magnitude at any point of the evolution and both decrease substantially during the timespan of observations. Therefore, the period decrease of V1309~Sco cannot be understood within a theory that predicts period evolution to have similar and constant timescales $\pdot/\ddot{P}$ and $P/\pdot$, such as the equilibrium tidal theory \citep{eggleton98,eggleton01}. However, \citet{eggleton12} asserted that the Darwin instability in V1309~Sco could act on a timescale as short as few years or even days. Furthermore, tidal dissipation efficiency can be significantly increased with non-linear effects such as resonances between the orbital motion and oscillations in the star, if they can be maintained long enough, for example by resonance locking.

The accelerating period decrease of V1309~Sco must be due to the onset of an instability that operates on a timescale much shorter than $\ttf$ yet longer than the dynamical timescale $P$ (since it took V1309~Sco several hundred days to reach optical maximum). This is compatible with dynamically unstable mass loss, which occurrs when a star tends to adiabatically expand with respect to the Roche lobe radius as a response to a removal of surface layers \citep[e.g.][]{paczynski72,plavec73,webbink77a,webbink77b,hjellming87,soberman97}.  {\citet{tylenda11}} also suggested that dynamically unstable mass loss operates in V1309~Sco and  I argue in favor of this possibility, because it allows to explain the unusually slow rise of luminosity of V1309~Sco.


\section{Connecting period change and luminosity increase}
\label{sec:wind}

After the binary variability disappeared, the luminosity of V1309~Sco slowly increased during $\sim 200$\,days as is shown in Figure~\ref{fig:lc}. This slow rise is longer than the dynamical time of the system, but comparable to the timescale on which the period of the binary orbit was changing just before the binary variability disappeared. In this Section, I describe how the period decay and slow luminosity increase can be connected using dynamical mass loss. The luminosity depends on the the mass loss rate from the system $\mout$, which is in turn proportional to the period change,
\beq
\frac{\pdot}{P} = -\mathcal{A} \frac{\mout}{M},
\label{eq:pdotp}
\eeq
where the mass outflow rate $\mout = -\dot{M}$ is assumed to be positive, $M=M_1+M_2$ is the total mass of the binary, and $\mathcal{A}$ is the constant of proportionality that depends on the mechanism of period change and mass loss.

Without detailed knowledge of the structure of V1309~Sco before the outburst, which can be eventually obtained from detailed modelling of the light curve, it is unknown whether the mass leaves the binary through the outer Lagrange point \citep{kuiper41,lubow75,shu79} or whether there is a combination of mass transfer and mass loss. I will discuss the case of mass loss through the outer Lagrange point, because it provides a self-contained relation between the period change and mass loss rate without additional parameters.

Mass loss through the outer Lagrange point will cause the orbital period to decrease. The exact form of relation between $\mout/M$ and $\pdot/P$ (value of $\mathcal{A}$ in Eq.~[\ref{eq:pdotp}]) depends on which star is actually losing the mass \citep{pribulla98}. I choose the case when the mass first flows from the primary to the envelope of the secondary and then is lost through the outer Lagrange point, which gives \citep[Eq.~(16)]{pribulla98}
\beq
\mathcal{A} = \frac{3(1+q)^2}{q}r(q)^2 - 3(1+q) + 1,
\label{eq:pdotpl2}
\eeq
where $r(q)$ is a non-dimensional slowly varying function of the mass ratio, $r(q)\approx 1.2$. The other cases discussed in \citet{pribulla98} differ by a factor of $\sim 2$ in $\mathcal{A}$. The order-of-magnitude estimate presented in this paper works for any $\mathcal{A} > 0$ since $\mathcal{A}$ is degenerate with other quantities such as opacity or velocity of the outflow. Given the estimate that $q\approx 0.1$ and $M\approx 1.7\,\msun$ from \citet{stepien11}, the period change of V1309~Sco implies $\mout \sim 2\times 10^{-4}\,\msun$\,yr$^{-1}$ at $t\approx 4300$.

For mass loss described by Equation~(\ref{eq:pdotp}), the orbital energy $E_{\rm orb} = -GM_1M_2/(2a)$ decays as
\beq
\dot{E}_{\rm orb} = E_{\rm orb}\left(\frac{2+3q-2\mathcal{A}}{3\mathcal{A}}  \right) \frac{\pdot}{P} \approx 1.5\times 10^{37}\frac{\rm ergs}{\rm s} \left(\frac{\pdot/P}{10^{-2}\,{\rm yr}^{-1}} \right),
\label{eq:eorb}
\eeq
where I assumed that $-\mout=\dot{M}=\dot{M}_1$. Parameters appropriate for V1309~Sco were used. The binary orbit loses energy, because $\mathcal{A} \gg 1$ for small $q$, and $E_{\rm orb}$ and $\pdot/P$ are both negative.

The kinematics of the matter leaving the outer Lagrange point was studied by \citet{shu79}. For $0.064 \leq q \leq 0.78$, the particles leaving L2 will fly to infinity in a ``garden sprinkler'' spiral pattern both in the co-rotating and inertial reference frames. Although the outflow starts as a stream originating at L2, it should surround the binary and become more isotropic as the binary rotates and the spiral arms spread both in the horizontal and vertical directions \citep{shu79}. I therefore assume that the outflow can be treated as spherically symmetric with radial density profile
\beq
\rho(r) \sim \frac{\mout}{4\pi r^2 \vesc},
\label{eq:mwind}
\eeq
where $\vesc$ is the characteristic velocity of the outflow, which I assume to be equal the escape velocity of the binary, $\vesc = \sqrt{2GM/a}$. 

Despite the rapid and accelerating period change, the system still evolves on a much longer timescale than the dynamical timescale ($P/\pdot \gg P$) and it is thus possible to assume quasistatic evolution. The power necessary to launch the outflow
\begin{eqnarray}
\dot{E}_{\rm out} &=& \frac{\mout \vesc^2}{2}  \approx \nonumber \\
&\approx& 2.9\times 10^{36}\,\frac{\rm ergs}{{\rm s}} \left( \frac{\mout}{10^{-4}\,\msun/{\rm yr}} \right)\left(\frac{\vesc}{300\,{\rm km}/{\rm s}}\right)^2.
\label{eq:ewind}
\end{eqnarray}
comes from the orbital decay. The energy decay of the binary orbit (Eq.~[\ref{eq:eorb}]) is sufficient to provide kinetic energy of the outflow with this high mass-loss rate.

I also assume that the outflow is isothermal with the color temperature of the binary, which should roughly hold until the material gets far from the binary \citep{shu79}. As the dynamical mass loss runs away, $\mout$ increases and the outflow eventually forms a photosphere with a radius
\begin{eqnarray}
\rtau(t) &\sim& \frac{\kappa \mout(t)}{4\pi \vesc} \sim \frac{\kappa M }{4\pi \vesc\mathcal{A}} \left|\frac{\pdot}{P}\right| \approx \nonumber \\
&\approx& 4.8 R_\odot \kappa_{-2} \left(\frac{M}{\msun}\right)\!\!\! \left( \frac{\vesc}{300\,{\rm km}/{\rm s}} \right)^{-1} \left|\frac{\pdot/P}{0.01\,{\rm yr}^{-1}}  \right|,
\label{eq:photo}
\end{eqnarray}
where $\kappa = \kappa_{-2} 10^{-2}$\,cm$^2$\,g$^{-1}$ is the characteristic opacity of the isothermal outflow. Although I assume that the outflow is nearly isothermal, a proper calculation of the outflow thermodynamic structure is necessary to precisely connect the mass loss rate with the position of the photosphere, because for the temperature range relevant for V1309~Sco, $\kappa$ grows steeply with temperature due to H$^-$ opacity \citep{alexander94,ferguson05} and the photosphere might be positioned at the largest gradient of opacity.

The binary optical radiation $L_{\rm binary}$ will be replaced by that from the photosphere of the outflow with luminosity
\beq
L_{\rm out} = 4\pi\rtau^2 \sigma T_{\rm out}^4,
\label{eq:l}
\eeq
where the photospheric temperature $T_{\rm out}$ should not be substantially different from the temperature of the binary $T_{\rm binary}$. The luminosity of the optically-thick outflow is thus dictated by the physics of the mass loss from the binary (Eq.~[\ref{eq:photo}]). The outflow should be radiatively inefficient in the sense that 
\beq
\eta = \frac{L_{\rm out}}{\dot{E}_{\rm out}} = \frac{\kappa^2 \mout \sigma T_{\rm out}^4}{2\pi \vesc^4} \ll 1,
\label{eq:eta}
\eeq
but no conversion of $\dot{E}_{\rm out}$ to $L_{\rm out}$ is necessary below the photosphere. Equivalently, the outflow thermal energy is much smaller than its kinetic energy.

\section{Mass loss rate evolution of V1309 Sco}
\label{sec:appl}

In this Section, I present a fit to the light curve and period evolution of V1309~Sco. The phenomenological prescriptions in the model are based on the prediction of Section~\ref{sec:wind} that $L \propto (\pdot/P)^2$ and previous works on dynamical mass loss. The values of the fit parameters are reasonable when interpreted within the scenario presented in Section~\ref{sec:wind}, which suggests that the presented model is consistent with the observations.

\begin{figure}
\centering
\plotone{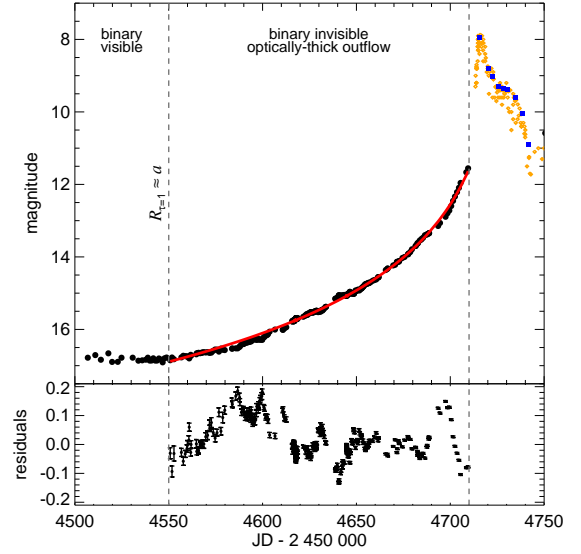}
\caption{Light curve of V1309~Sco during the slow rise to maximum constructed from OGLE $I$-band data (gray points), AAVSO visual and $V$-band data (open orange diamonds), and ASAS $V$-band \citep[blue squares,][]{pojmanski97} is shown in the top panel. Vertical dashed lines mark moments of complete obscuration of the binary ($t \approx 4550$) and the beginning of rapid brightening ($t = 4710$). The red line is the fit of the optically-thick outflow model to the data (Eq.~[\ref{eq:photo}]). The residuals of the fit in the bottom panel show changes correlated on the timescale of $10$ to $20$ days with amplitude higher than photometric uncertainties.}
\label{fig:lc}
\end{figure}

Figure~\ref{fig:lc} shows the slow rise to optical maximum of V1309~Sco ($4550 \leq t \leq 4710$), which I explain as the expanding photosphere of an optically-thick outflow powered by the dynamical mass loss. The orbital variations disappeared between $t\approx 4300$ and $4550$, which constrains $\kappa$ to be between $5\times 10^{-3}$ and $1.3\times 10^{-2}$\,cm$^2$\,g$^{-1}$ ($\rtau \approx a$, Eq.~[\ref{eq:photo}]). This is a reasonable result for a relatively low temperature material \citep{alexander94,ferguson05}. I assume that $\rtau = a$ at $t\approx 4550$, which gives $\kappa \approx 5\times 10^{-3}$\,cm$^2$\,g$^{-1}$. The photospheric temperature was not measured during the optically-thick outflow phase, but the low opacity suggests photospheric temperatures $\lesssim 7000$\,K. Before the binary became obscured, \citet{tylenda11} found $T_{\rm binary} \approx 4500$\,K, while at the optical maximum the temperature was between about $5000$ and $6000$\,K \citep{mason10,tylenda11}. At optical maximum, the FWHM of the Balmer lines was $150$\,km\,s$^{-1}$ \citep{mason10}, consistent with the $\vesc = 300$\,km\,s$^{-1}$ assumed here.

By combining the directly observed orbital period change and the slow brightening due to the optically-thick outflow, it is possible to constrain the total mass loss evolution of the primary star in V1309~Sco. I first focus on the slow brightening. To predict the change in luminosity as a function of time, the model of Section~\ref{sec:wind} requires the time evolution of the mass loss rate. Following \citet{paczynski72} and \citet{webbink77b}, I assume that the mass loss rate runs away as a power law
\beq
\frac{\mout }{M} = \frac{\gamma}{|t-t_0|^\delta},
\label{eq:mdot}
\eeq
with a singularity at $t_0$. Combining Equations~(\ref{eq:photo}) and (\ref{eq:mdot}), the observed magnitude is
\beq
m(t) = 5\delta\log|t-t_0|+C,
\label{eq:lc_fit}
\eeq
where $C$ hides all remaining parameters (including $\gamma$, distance modulus and extinction), which I assume to be constant with $t$. The optically-thick outflow phase shown in Figure~\ref{fig:lc} gives $t_0 = 4730 \pm 0.5$ and $\delta = 1.12 \pm 0.01$ and the model matches the slow brightening remarkably well with only three parameters $t_0$, $\delta$, and $C$. The fit is valid only until $t\approx 4710$, because at that point the brightening of V1309~Sco significantly accelerated and proceeded on approximately the dynamical timescale of the system. It is thus not surprising that the fit gives $t_0$ about $20$\,days later than the start of the dynamic brightening.

The parameter $\delta$ has a clear physical meaning directly related to the properties of the surface layer of the star experiencing mass loss. Specifically, $\delta = (2n+3)/(2n+1)$, where $n$ is the polytropic index \citep{jedrzejec69,paczynski72,webbink77b}, which gives $\delta=3/2$ and $9/7$ for a fully convective and radiative star, respectively. The value obtained from fitting V1309~Sco implies very high values of $n$ and nearly isothermal surface layer during the dynamical mass loss. The model presented here can in principle observationally constrain the behavior of stars in a response to large amounts of mass loss, which has been calculated only theoretically \citep{woods11,passy12}. However, a more detailed model of the outflow structure is necessary to provide a robust value of $\delta$, because the estimate of $\delta$ can be biased if quantities such as $\kappa$ and the outflow velocity smoothly evolve in time.


\begin{figure}
\centering
\plotone{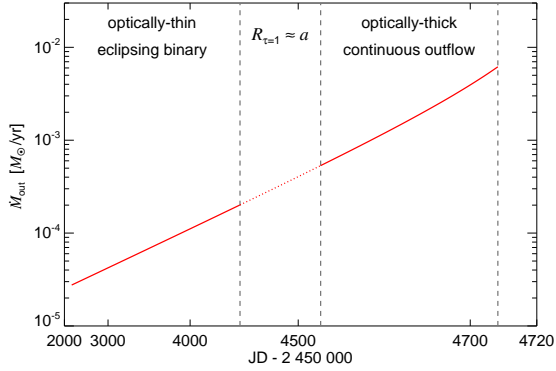}
\caption{Total mass loss rate of V1309~Sco reconstructed by combining period measurements (Fig.~\ref{fig:period}) and the slow brightening (Fig.~\ref{fig:lc}) assuming $q=0.1$, which gives $\mathcal{A} \approx 50$.} 
\label{fig:mdot}
\end{figure}

The fit of the slow brightness rise of V1309~Sco determines well $\delta$ and $t_0$, but cannot constrain the absolute scale of the mass loss rate, because $\gamma$ enters only through the additive constant $C$, which depends also on uncertain values of distance, extinction, $\kappa$, $\vesc$, and others. Fortunately, the constraints from the directly observed $P(t)$ are complementary to the slow brightness rise. The observed $P(t)$ cannot uniquely determine $t_0$, $\gamma$, and $\delta$, because the binary becomes obscured too far from the singularity at $t_0$. But with $\delta$ and $t_0$ constrained by the slow brightness rise, $P(t)$ sets the absolute scale of the mass loss rate $\gamma$ (assuming $\mathcal{A}(q)$ is known). The explicit period evolution can be obtained by integrating Equations~(\ref{eq:pdotp}) and (\ref{eq:mdot}), which gives
\beq
P(t) = P_0 \exp \left[\frac{\mathcal{A}\gamma}{1-\delta} |t-t_0|^{1-\delta}  \right],
\label{eq:period_fit}
\eeq
where $P_0$ is the integration constant and the Equation is valid for $\delta \ne 1$. Applying this to the directly observed period evolution with $\delta$ and $t_0$ fixed to the results from the light curve fit gives values of $P_0$ and $\gamma$. In particular, the uncertainty in $\gamma$ due to the scatter in $P(t)$ is $\sim 5\%$ if there is no uncertainty in $\mathcal{A}$. It is worth noting that although $\delta$ and $t_0$ are fixed during period fitting, their values cannot be chosen arbitrarily to match the period change. For example, with $\delta \approx 2.25$ the curvature of $P(t)$ can be matched only with $t_0$ about $1000$\,days after the optical maximum. This provides an additional consistency check on the model.

Figure~\ref{fig:mdot} shows the reconstructed evolution of $\mout$ assuming $q=0.1$, which gives $\mathcal{A} \approx 50$. The quasistatic evolution lasted $\sim 2400$\,days during which $\mout$ increased by more than two orders of magnitude. At $t\approx 4710$, the binary orbital period is estimated to be $P \approx 1.387$\,days, which translates to $97.7\%$ of the separation at the beginning of observations. The total mass lost from the binary is $\sim 1.2\times 10^{-3}\,\msun$ with a kinetic energy of $\sim 1.1\times 10^{45}$\,ergs. The total energy radiated during the optically-thick outflow phase is much smaller, $\sim 1.2\times 10^{43}$\,ergs. The wind is radiatively inefficient, $\eta \simeq 1\%$, with peak efficiency of $\simeq 5\%$ at $t=4710$ assuming $T=5000$\,K. Since $\mathcal{A}$ is a relatively steep at small $q$, the values of $\mout$ are somewhat sensitive to the assumed mass ratio of the binary. For example, changing $q$ to $0.05$ or $0.2$ gives $\mathcal{A}$ of about $93$ or $29$, respectively.

The bottom panel of Figure~\ref{fig:lc} shows the residuals with respect to the fit during the slow rise to the maximum. The residuals are not randomly distributed around zero, but are correlated on a timescale of a few to tens of days with amplitude greater than the photometric uncertainty. The most likely explanation is that these variations are caused by the non-isotropic distribution of matter in the outflow, a possible sign of the ``garden-sprinkler'' spiral created by the mass loss from the outer Lagrange point \citep{shu79}. If the outflowing material keeps its tangential velocity, the characteristic duration of these correlated changes should increase with time as the increasing $\mout$ moves the photosphere to larger radii. The separation of the first two prominent peaks at $t\approx 4588$ and $4600$ is about $12$\,days, while the last and second-to-last peaks are separated by about $20$\,days, however, the variations in between are more stochastic and it is hard to deduce any timescale. The timescale of variations relative to $P$ implies photospheric radii similar to what is required to explain the observed slow brightening.

V1309~Sco likely did not become optically thick from all angles simultaneously. \citet{tylenda11} showed how the shape of the eclipsing binary light curve evolved as a function of time: the magnitude difference between the maximum at phase $0.75$ and minima at phases $0.0$ and $0.5$ decreases to essentially zero with increasing time, while the maximum at phase $0.25$ appears to be relatively unchanged. At phases $0.25$ and $0.75$, we see both stars from the side with observed surface area maximized. The light curve of V1309~Sco can be explained by a gradually growing ``dark spot'' on one of the stars that has maximum visible surface area around phase $0.75$, and has little effect on the flux at other phases. Putting a ``bright spot'' on the stars fully visible at either phase $0.25$ or $0.75$ would lead to an increase of the total amplitude of variability, because the binary would get brighter at these phases relative to phases $0.0$ and $0.5$, which is not observed. If the mass loss occurs as a gradually spreading stream originating from L2, the density and hence optical depth of the material will be higher on one side of the binary than the other and could explain the evolution of the binary light curve. The material would have to be colder than the stars to cause obscuration (and possibly the drop in luminosity when the binary variability disappeared), while later causing the brightening due to large surface area. Detailed modelling of the pre-outburst light curve of V1309~Sco could thus constrain the spatial structure of the mass loss.


\section{Discussions}
\label{sec:disc}

The model presented in Section~\ref{sec:wind} connects together the directly observed period changes and the subsequent slow brightness rise through intense mass loss. However, the quasistatic evolution of the outflow photosphere is followed by a rapid brightening that started on $t\approx 4710$ and lasted $\sim 5$\,days, which is only a few orbits. The short timescale implies that a different physical effect quenches the quasistatic evolution of the optically-thick outflow. 

The rapid brightening can be achieved by significantly increasing the radiation efficiency in the wind for example due to a passing shock wave; converting $\sim 70\%$ of wind power at $t\approx 4710$ to radiation is enough to explain the peak luminosity derived by \citet{tylenda11}. However, it is not clear what would be the origin of such a shock. Instead, the rapid brightening might be a sign of the merger of the underlying binary. \citet{ivanova13a} and \citet{nandez13} performed simulations of a binary merger with parameters appropriate for V1309~Sco and found that the merger ejects a shell of several $10^{-2}\,\msun$ on the dynamical timescale and the rapid brightening is powered by the associated recombination energy. Their results well match the properties of V1309~Sco around the maximum brightness, after $t\approx 4710$.

\subsection{Implications for similar transients}
\label{sec:implications}

Could the model presented here be used to predict when the rapid brightening occurs? Such a prediction would be extremely interesting from an observational point of view, but is difficult based on order-of-magnitude estimates alone. For example, the decrease of the semi-major axis between $t\approx 2300$ and $4710$ inferred from period changes and optically-thick outflow is less than $3\%$, so any condition based on $a$ would require signficant fine tuning.

Instead, it is interesting to ask what is the range of validity of the model presented here and how would the optically-thick outflow look like in binaries with different properties, especially the orbital period. First, in order to get an optically-thick outflow, the mass loss rate must be sufficiently high to produce a photosphere at $\rtau \gtrsim a$. Apart from the opacity of the material, this condition depends on the characteristic velocity of the outflow $\vesc$, which scales with the binary period (Eq.~[\ref{eq:photo}]). As a result, the critical $\mout$, which gives $\rtau = a$, scales as $\mout \propto a^{1/2} \propto P^{1/3}$ for constant opacity. 


Second, it is also reasonable to assume that the optically-thick outflow does not radiate any substantial part of its kinetic energy. In the formalism of Equation~(\ref{eq:eta}), $\eta \ll 1$. I choose a rather arbitrary maximum value of $\eta = 0.05$, which corresponds to the inferred properties of the outflow of V1309~Sco at $t\approx 4710$. Lines of constant $\eta$ imply scaling $\mout \propto a^{-2} \propto P^{-4/3}$ for constant opacity. 

\begin{figure}
\plotone{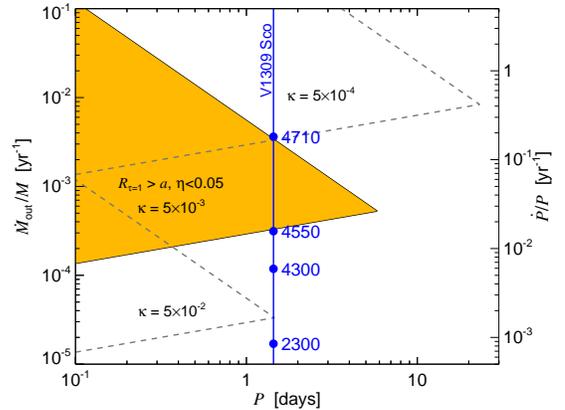}
\centering
\caption{
Constraints on the properties of a continuous outflow with mass loss rate $\mout$ from binaries with orbital period $P$. The mass was assumed to be $M=1.7\,\msun$ and the mass loss rate is linked to period change $\pdot$ using Eqs.~(\ref{eq:pdotp}) and (\ref{eq:pdotpl2}) with $q=0.1$. The dark yellow wedge shows the allowed parameter space for the continuous optically-thick outflow, where slow brightening can be observed. The lower limit of the wedge is due to requirement $\rtau > a$ (Eq.~[\ref{eq:photo}]) and the upper limit is given by the luminosity constraint $\eta < 0.05$ (Eq.~[\ref{eq:eta}]), both assuming opacity $\kappa = 5\times 10^{-3}$\,cm$^2$\,g$^{-1}$ appropriate for V1309~Sco and consistent with $T\sim 5000$\,K material \citep{alexander94,ferguson05}. Two additional wedges drawn with grey dashed lines are similar constraints but for opacities lower and higher by a factor of $10$. 
The blue vertical line shows the orbital period of V1309~Sco and the blue dots show the derived mass loss rates at specific times (see also Figs.~\ref{fig:lc} and \ref{fig:mdot}).
}
\label{fig:4}
\end{figure}

These constraints are summarized in Figure~\ref{fig:4} in the parameter space of the relative outflow rate $\mout$ and orbital period $P$. The yellow wedge shows the combined constraint $\rtau>a$ and $\eta < 0.05$ for opacity appropriate for V1309~Sco and similar constraints for higher and lower opacities are shown with grey dashed lines. These wedges indicate that the behavior similar to V1309~Sco (slow luminosity increase due to an optically-thick outflow) can be expected only in binaries with $P \lesssim 30$\,days, which produce relatively cool and thus not very opaque outflows. Although high opacities require only small $\mout$, the outflow luminosity would be very much higher than its kinetic energy. Similarly, long-period binaries would require very low $\kappa$ and thus very high $\mout$. 

What would be the observable appearance of binaries that are losing mass with high $\mout$ but increasing the binary brightness would require unreasonably high $\eta$? The outflow in such binaries would neccessarily produce high optical depths $\tau$, but the outflow could only reprocess the binary luminosity to lower temperatures, $T_{\rm eff} \approx \tau^{-1/4} T_{\rm binary}$, keeping the total luminosity essentially fixed. Low temperatures in such outflows would be very conducive for dust formation and such binaries would appear as very cool and dusty stars. If the mass loss was due to dynamical non-conservative mass transfer that lasted only a short amount of time, the result would be an expanding cold dusty shell. In either case, such obscured objects could potentially be identified by infrared observations.

A recent event OGLE-2002-BLG-360 \citep{tylenda13} showed brightening on a timescale about $\sim 3.5$ times longer than V1309~Sco. The lack of any significant short-period pre-outburst variation is in line with an optically-thick outflow during dynamical mass loss and a reasonable fit to the rising part of the light curve is possible with Equations~(\ref{eq:photo}--\ref{eq:mdot}). Unlike V1309~Sco, OGLE-2002-BLG-360 did not show a rapid brightening on a dynamical timescale. Perhaps in this case, the dynamical mass loss was due to nonconservative mass transfer, which did not end with a merger of the two stars. The slow rise to maximum distinguishes V1309~Sco, OGLE-2002-BLG-360, and potentially also V4332~Sgr \citep{kimeswenger06,martini99} from other ``red novae'' such as V838~Mon and M31~RV, which seem to show only rapid brightening from a quiescent magnitude \citep[e.g.][]{sharov93,munari02,sobotka02}. It is also worth noting that these objects differ from M85 OT2006-1, SN 2008S, and NGC300 OT2008-1 in their spectral evolution, luminosity, and progenitors \citep{kulkarni07,prieto08,berger09,bond09,smith09,thompson09}. The non-detection of a slow rise to maximum might simply be due to a lack of pre-outburst measurements: there would be no indication of a slow rise in V1309~Sco without the extensive OGLE data. It is worth mentioning that the tentative category of ``red novae'' is heterogenous in many other aspects as well. Transients due to dynamical mass loss might also be confused with other sources of variability such as microlensing, which was the case of OGLE-2002-BLG-360 \citep{tylenda13}.

\section{Conclusions}
\label{sec:conc}

I focused on two previously overlooked features of the outburst of V1309~Sco: (1) the rapidly accelerating pre-outburst period decay (Fig.~\ref{fig:period}) and (2) the slow rise to optical maximum (Fig.~\ref{fig:lc}). I showed how these two observed features can be unified through a model of a continuous optically-thick outflow generated by dynamical but long-lasting mass loss. I obtain remarkably good fits of the light curve and period evolution (Figs.~\ref{fig:period} and \ref{fig:lc}) and interpret the fitted parameters within the optically-thick outflow model. In particular,  I constrain the mass loss rate of the binary (Fig.~\ref{fig:mdot}). The model assumes that most of the orbital decay energy goes into the kinetic energy of the outflow.

This new class of transients distinguished by a slow rise to optical maximum that are driven by dynamical mass loss deserves much more detailed and realistic calculation. It is especially important to properly calculate the thermodynamic structure of the wind, including realistic opacities and dust formation, which occurred in OGLE-2002-BLG-360 \citep{tylenda13} and after optical maximum in V1309~Sco \citep{nicholls13}. The optically-thick outflow is also likely asymmetric, which is possibly observed in the residuals of the fit (Fig.~\ref{fig:lc}). Proceeding forward with greater realism will be rewarding, because the optically-thick outflow encodes the mass-loss evolution of the underlying binary and is potentially important for other similar types of systems involving massive stars \citep[e.g.][]{belczynski08,eldridge08,demink13}, white dwarfs \citep[e.g.][]{webbink84,verbunt88,benz90,dan11}, and neutron stars \citep[e.g.][]{rasio94,rosswog05,dsouza06}.

\acknowledgements

I am grateful to T.~Thompson and C.~Kochanek for encouragement, discussions, suggestions, and a detailed reading of the manuscript. I thank K.~Stanek for encouragement and suggestions, and to R.~Tylenda for providing me with period measurements of V1309~Sco. I thank the referees for useful and constructive suggestions, J.~Goodman for discussions and R.~Rafikov for suggesting the tidal resonance. Support for Program number HST-HF-51327.01 was provided by NASA through a Hubble Fellowship grant from the Space Telescope Science Institute, which is 
operated by the Association of Universities for Research in Astronomy, Incorporated, under NASA contract NAS5-26555. I thank the OGLE team for making their data public. I acknowledge with thanks the variable star observations from the AAVSO International Database contributed by observers worldwide and used in this research.

{\it Facility:} \facility{AAVSO}


\begin{thebibliography}{}
\bibitem[Alexander \& Ferguson(1994)]{alexander94} Alexander, D.~R., \& Ferguson, J.~W.\ 1994, \apj, 437, 879
\bibitem[Belczynski et al.(2008)]{belczynski08} Belczynski, K., Kalogera, V., Rasio, F.~A., et al.\ 2008, \apjs, 174, 223  
\bibitem[Benz et al.(1990)]{benz90} Benz, W., Cameron, A.~G.~W., Press, W.~H., \& Bowers, R.~L.\ 1990, \apj, 348, 647
\bibitem[Berger et al.(2009)]{berger09} Berger, E., Soderberg, A.~M., Chevalier, R.~A., et al.\ 2009, \apj, 699, 1850
\bibitem[Bond et al.(2009)]{bond09} Bond, H.~E., Bedin, L.~R., Bonanos, A.~Z., et al.\ 2009, \apjl, 695, L154    
\bibitem[Dan et al.(2011)]{dan11} Dan, M., Rosswog, S., Guillochon, J., \& Ramirez-Ruiz, E.\ 2011, \apj, 737, 89 
\bibitem[de Mink et al.(2013)]{demink13} de Mink, S.~E., Langer, N., Izzard, R.~G., Sana, H., \& de Koter, A.\ 2013, \apj, 764, 166
\bibitem[D'Souza et al.(2006)]{dsouza06} D'Souza, M.~C.~R., Motl, P.~M., Tohline, J.~E., \& Frank, J.\ 2006, \apj, 643, 381  
\bibitem[Eggleton(1983)]{eggleton83} Eggleton, P.~P.\ 1983, \apj, 268, 368
\bibitem[Eggleton et al.(1998)]{eggleton98} Eggleton, P.~P., Kiseleva, L.~G., \& Hut, P.\ 1998, \apj, 499, 853 
\bibitem[Eggleton \& Kiseleva-Eggleton(2001)]{eggleton01} Eggleton, P.~P., \& Kiseleva-Eggleton, L.\ 2001, \apj, 562, 1012
\bibitem[Eggleton(2012)]{eggleton12} Eggleton, P.~P.\ 2012, Journal of Astronomy and Space Sciences, 29, 145 
\bibitem[Eldridge et al.(2008)]{eldridge08} Eldridge, J.~J., Izzard, R.~G., \& Tout, C.~A.\ 2008, \mnras, 384, 1109 
\bibitem[Ferguson et al.(2005)]{ferguson05} Ferguson, J.~W., Alexander, D.~R., Allard, F., et al.\ 2005, \apj, 623, 585 
\bibitem[Hjellming \& Webbink(1987)]{hjellming87} Hjellming, M.~S., \& Webbink, R.~F.\ 1987, \apj, 318, 794 
\bibitem[Iben \& Livio(1993)]{iben93} Iben, I., Jr., \& Livio, M.\ 1993, \pasp, 105, 1373 
\bibitem[Ivanova et al.(2013a)]{ivanova13a} Ivanova, N., Justham, S., Avendano Nandez, J.~L., \& Lombardi, J.~C.\ 2013a, Science, 339, 433
\bibitem[Ivanova et al.(2013b)]{ivanova13b} Ivanova, N., Justham, S., Chen, X., et al.\ 2013b, \aapr, 21, 59
\bibitem[Jedrzejec(1969)]{jedrzejec69} Jedrzejec, E.\ 1969, M.S.\ thesis, Warsaw University
\bibitem[Kimeswenger(2006)]{kimeswenger06} Kimeswenger, S.\ 2006, Astronomische Nachrichten, 327, 44
\bibitem[Kuiper(1941)]{kuiper41} Kuiper, G.~P.\ 1941, \apj, 93, 133  
\bibitem[Kulkarni et al.(2007)]{kulkarni07} Kulkarni, S.~R., Ofek, E.~O., Rau, A., et al.\ 2007, \nat, 447, 458 
\bibitem[Livio \& Soker(1988)]{livio88} Livio, M., \& Soker, N.\ 1988, \apj, 329, 764 
\bibitem[Lubow \& Shu(1975)]{lubow75} Lubow, S.~H., \& Shu, F.~H.\ 1975, \apj, 198, 383 
\bibitem[Martini et al.(1999)]{martini99} Martini, P., Wagner, R.~M., Tomaney, A., et al.\ 1999, \aj, 118, 1034 
\bibitem[Mason et al.(2010)]{mason10} Mason, E., Diaz, M., Williams, R.~E., Preston, G., \& Bensby, T.\ 2010, \aap, 516, A108
\bibitem[Munari et al.(2002)]{munari02} Munari, U., Henden, A., Kiyota, S., et al.\ 2002, \aap, 389, L51 
\bibitem[Nandez et al.(2013)]{nandez13} Nandez, J.~L.~A., Ivanova, N., \& Lombardi, J., Jr 2013, arXiv:1311.6522 
\bibitem[Nicholls et al.(2013)]{nicholls13} Nicholls, C.~P., Melis, C., Soszy{\'n}ski, I., et al.\ 2013, \mnras, 431, L33 
\bibitem[Paczy{\'n}ski \& Sienkiewicz(1972)]{paczynski72} Paczy{\'n}ski, B., \& Sienkiewicz, R.\ 1972, \actaa, 22, 73 
\bibitem[Paczy{\'n}ski(1976)]{paczynski76} Paczy{\'n}ski, B.\ 1976, Structure and Evolution of Close Binary Systems, 73, 75
\bibitem[Passy et al.(2012)]{passy12} Passy, J.-C., Herwig, F., \& Paxton, B.\ 2012, \apj, 760, 90 
\bibitem[Plavec et al.(1973)]{plavec73} Plavec, M., Ulrich, R.~K., \& Polidan, R.~S.\ 1973, \pasp, 85, 769 
\bibitem[Pojmanski(1997)]{pojmanski97} Pojmanski, G.\ 1997, \actaa, 47, 467
\bibitem[Pribulla(1998)]{pribulla98} Pribulla, T.\ 1998, Contributions of the Astronomical Observatory Skalnate Pleso, 28, 101
\bibitem[Prieto et al.(2008)]{prieto08} Prieto, J.~L., Kistler, M.~D., Thompson, T.~A., et al.\ 2008, \apjl, 681, L9   
\bibitem[Rasio \& Shapiro(1994)]{rasio94} Rasio, F.~A., \& Shapiro, S.~L.\ 1994, \apj, 432, 242
\bibitem[Rosswog(2005)]{rosswog05} Rosswog, S.\ 2005, \apj, 634, 1202  
\bibitem[Sharov(1993)]{sharov93} Sharov, A.\ 1993, Astronomy Letters, 19, 83 
\bibitem[Shu et al.(1979)]{shu79} Shu, F.~H., Anderson, L., \& Lubow, S.~H.\ 1979, \apj, 229, 223
\bibitem[Smith et al.(2009)]{smith09} Smith, N., Ganeshalingam, M., Chornock, R., et al.\ 2009, \apjl, 697, L49  
\bibitem[Soberman et al.(1997)]{soberman97} Soberman, G.~E., Phinney, E.~S., \& van den Heuvel, E.~P.~J.\ 1997, \aap, 327, 620 
\bibitem[Sobotka et al.(2002)]{sobotka02} Sobotka, P., \v{S}melcer, L., Pejcha, O., et al.\ 2002, Information Bulletin on Variable Stars, 5336, 1
\bibitem[Soker \& Tylenda(2003)]{soker03} Soker, N., \& Tylenda, R.\ 2003, \apjl, 582, L105 
\bibitem[Soker \& Tylenda(2006)]{soker06} Soker, N., \& Tylenda, R.\ 2006, \mnras, 373, 733 
\bibitem[St{\c e}pie{\'n}(2011)]{stepien11} St{\c e}pie{\'n}, K.\ 2011, \aap, 531, A18 
\bibitem[Taam \& Sandquist(2000)]{taam00} Taam, R.~E., \& Sandquist, E.~L.\ 2000, \araa, 38, 113 
\bibitem[Thompson et al.(2009)]{thompson09} Thompson, T.~A., Prieto, J.~L., Stanek, K.~Z., et al.\ 2009, \apj, 705, 1364 
\bibitem[Tylenda \& Soker(2006)]{tylenda06} Tylenda, R., \& Soker, N.\ 2006, \aap, 451, 223 
\bibitem[Tylenda et al.(2011)]{tylenda11} Tylenda, R., Hajduk, M., Kami{\'n}ski, T., et al.\ 2011, \aap, 528, A114
\bibitem[Tylenda et al.(2013)]{tylenda13} Tylenda, R., Kaminski, T., Udalski, A., et al.\ 2013, arXiv:1304.1694
\bibitem[Verbunt \& Rappaport(1988)]{verbunt88} Verbunt, F., \& Rappaport, S.\ 1988, \apj, 332, 193 
\bibitem[Webbink(1977a)]{webbink77a} Webbink, R.~F.\ 1977a, \apj, 211, 486
\bibitem[Webbink(1977b)]{webbink77b} Webbink, R.~F.\ 1977b, \apj, 211, 881  
\bibitem[Webbink(1984)]{webbink84} Webbink, R.~F.\ 1984, \apj, 277, 355  
\bibitem[Woods \& Ivanova(2011)]{woods11} Woods, T.~E., \& Ivanova, N.\ 2011, \apjl, 739, L48 
\end{thebibliography}
\end{document}